\documentclass[aps,pra,twocolumn,superscriptaddress, nofootinbib]{revtex4-1}
\pdfoutput=1

\usepackage[utf8]{inputenc}
\usepackage[english]{babel}
\usepackage[T1]{fontenc}
\usepackage{amsmath}
\usepackage{hyperref}
\usepackage{tikz}
\usepackage{lipsum}
\usepackage{float}
\usepackage{graphicx}
\usepackage{latexsym}
\usepackage{amssymb}
\usepackage{csquotes}
\usepackage{nicefrac}
\usepackage{amsfonts}
\usepackage{enumerate}
\usepackage{upgreek}
\usepackage{subfigure}
\usepackage{placeins}
\usepackage{bm}
\usepackage[version=3]{mhchem}
\usepackage{nicefrac}
\usepackage{bbold}
\usepackage{verbatim}
\usepackage{multirow}
\usepackage{color}
\usepackage{comment}
\usepackage{xcolor}
\usepackage{soul}
\usepackage{bibunits}

\usepackage{braket}

\newcommand{\C}{\hat{c}}

\renewcommand{\vec}[1]{\mathbf{#1}}

\begin{document}

\title{Simulating the two-dimensional $t$-$J$ model at finite doping\\ with neural quantum states}

\author{Hannah Lange }
\thanks{These authors contributed equally. $ \quad \quad \quad\quad \quad\quad\quad \quad\quad \quad\quad \quad$ \texttt{hannah.lange@physik.uni-muenchen.de \\ annika.boehler@physik.uni-muenchen.de}}
\affiliation{Department of Physics and Arnold Sommerfeld Center for Theoretical Physics (ASC), Ludwig-Maximilians-Universit\"at M\"unchen, Theresienstr. 37, M\"unchen D-80333, Germany}
\affiliation{Munich Center for Quantum Science and Technology, Schellingstr. 4, Munich D-80799, Germany}
\affiliation{Max-Planck-Institute for Quantum Optics, Hans-Kopfermann-Str.1, Garching D-85748, Germany}

\author{Annika Böhler }
\thanks{These authors contributed equally. $ \quad \quad \quad\quad \quad\quad\quad \quad\quad \quad\quad \quad$ \texttt{hannah.lange@physik.uni-muenchen.de \\ annika.boehler@physik.uni-muenchen.de}}
\affiliation{Department of Physics and Arnold Sommerfeld Center for Theoretical Physics (ASC), Ludwig-Maximilians-Universit\"at M\"unchen, Theresienstr. 37, M\"unchen D-80333, Germany}
\affiliation{Munich Center for Quantum Science and Technology, Schellingstr. 4, Munich D-80799, Germany}

\author{Christopher Roth}
\affiliation{Center for Computational Quantum Physics, Flatiron Institute, 162 5th Avenue, New York, NY 10010, USA}

\author{Annabelle Bohrdt}
\affiliation{University of Regensburg, Universitätsstr. 31, Regensburg D-93053, Germany}
\affiliation{Munich Center for Quantum Science and Technology, Schellingstr. 4, Munich D-80799, Germany}

\date{\today}
\begin{abstract}

Simulating large, strongly interacting fermionic systems remains a major challenge for existing numerical methods. In this work, we introduce Gutzwiller projected hidden fermion determinant states (G-HFDS) to simulate the strongly interacting limit of the Fermi-Hubbard model, namely the $t$-$J$ model, across the entire doping regime. We demonstrate that the G-HFDS achieve energies competitive with matrix product states (MPS) on lattices as large as $10 \times 10$ sites while using several orders of magnitude fewer parameters, suggesting the potential for efficient application to even larger system sizes. This remarkable efficiency enables us to probe low-energy physics across the full doping range, providing new insights into the competition between kinetic and magnetic interactions and the nature of emergent quasiparticles. Starting from the low-doping regime, where magnetic polarons dominate the low energy physics, we track their evolution with increasing doping and different next-nearest neighbor hopping amplitudes through analyses of spin and polaron correlation functions as well as the Fermi surface. Our findings demonstrate the potential of determinant-based neural quantum states with inherent fermionic sign structure, opening the way for simulating large-scale fermionic systems at any particle filling.
\end{abstract}
\maketitle

Understanding the microscopic physics of large, interacting quantum materials such as unconventional superconductors \cite{Bednorz1986} has been a long-standing challenge in theoretical condensed matter physics. A model that is believed to be relevant to many phases observed in cuprate superconductors is the two-dimensional Fermi-Hubbard model -- or its high interaction limit, the $t$-$J$ model,
\begin{align}
        \mathcal{H}_{t-J}=&-t \sum_{\langle \vec{i},\vec{j}\rangle, \sigma}\mathcal{P}_G\left( \hat{c}^{\dagger}_{\vec{i},\sigma} \hat{c}_{\vec{j},\sigma}+\mathrm{h.c.}\right) \mathcal{P}_G\notag  \\
        &-t^\prime \sum_{\langle\langle  \vec{i},\vec{j}\rangle\rangle, \sigma}\mathcal{P}_G\left( \hat{c}^{\dagger}_{\vec{i},\sigma} \hat{c}_{\vec{j},\sigma}+\mathrm{h.c.}\right) \mathcal{P}_G\notag  \\&+J\sum_{\langle\vec{i},\vec{j}\rangle}\left(\hat{S}_{\vec{i}}\cdot \hat{S}_{\vec{j}}-\frac{1}{4}  \hat{n}_{\vec{i}}\hat{n}_{\vec{j}}\right) 
        \label{eq:tJ}
\end{align}
with the fermionic annihilation (creation) operators $\hat{c}^{(\dagger)}_{\vec{i},\sigma}$ at site $\vec{i}$ with spin $\sigma$, spin operators  $\hat{\vec{S}}_{\vec{i}}=\sum_{\sigma,\sigma^\prime}\hat{c}^{\dagger}_{\vec{i},\sigma} \frac{\vec{\sigma}_{\sigma \sigma^\prime}}{2}\hat{c}_{\vec{i},\sigma^\prime} $ and density operators $\hat{n}_{\vec{i}}$. Furthermore, $\mathcal{P}_G$ projects out states with more than one particle per site. At half filling, the system realizes an antiferromagnetic (AFM) Mott insulator. The competition between kinetic (next-) nearest neighbor hopping amplitude $t^{(\prime)}$ and antiferromagnetic energy scale $J$ leads to the formation of magnetic polarons, i.e. heavily dressed dopants  \cite{Bulaevski1968,Brinkman1970,Sachdev1989,Kane1989,Grusdt2018}, at low hole doping, and is believed to give rise to the pseudogap, superconductivity and stripe order upon further doping \cite{OMahony2022,Badoux2016, Jiang2021}. Finally, the system becomes a Fermi liquid at high hole doping \cite{Keimer2015}. The extent of these phases depends on the type of dopants, where hole and particle dopings are modeled by different signs of $t^\prime$. \\

Despite its seeming simplicity, simulating the Fermi-Hubbard or the $t$-$J$ model, particularly at finite hole doping, has proven to be highly challenging, even with impressive numerical studies \cite{Qin2020,Schaefer2021,xu2023,Arovas2022} and recent advances on quantum simulation platforms \cite{Bohrdt2021}. While variational methods like matrix product states (MPS) \cite{SCHOLLWOCK201196} become prohibitively expensive when applied to large two-dimensional systems, other established techniques like Quantum Monte Carlo (QMC) \cite{Becca2017} face the sign problem at finite doping, which can be overcome by auxiliary field QMC with the cost of introducing a bias \cite{xu2023}. Neural quantum states (NQS) introduced by Carleo and Troyer \cite{Carleo2017} can potentially overcome some of these limitations: 
NQS have demonstrated the ability to capture states with volume-law entanglement \cite{Sharir2022,Deng2017,Gao2017,denis2023comment,Levine2019} and have been shown to outperform traditional methods like MPS in two-dimensional spin systems \cite{Rende2024,chen2023efficient,Reh2023,wang2023variational,roth2023high}. While early research predominantly focused on spin-$1/2$ systems, recent efforts have increasingly turned to the simulation of bosonic and fermionic systems \cite{lange2024architecturesapplicationsreviewneural,medvidović2024neuralnetwork,nomura2024quantummanybodysolverusing}.\\

Here, we employ a neural network approach to simulate the two-dimensional $t$-$J$ model, namely Gutwiller projected \textit{hidden fermion determinant states} (G-HFDS). We show that relative to MPS wavefunctions, the G-HFDS are extremely efficient,  particularly at intermediate doping. Specifically, G-HFDS can achieve the same energies as MPS, using only a very small fraction of the parameters. As a consequence of this efficiency, we are able to study the physical properties of the $t$-$J$ model in a full doping scan: We calculate spin-spin and polaron correlations, compare them to experimental observations \cite{Koepsell2021,Cheuk2016}, and track their evolution with different next-nearest neighbor hopping amplitudes, corresponding to hole and particle doping. Furthermore, G-HFDS allow for periodic boundary conditions, enabling the evaluation of momentum resolved quantities. Specifically we analyze the momentum resolved density -- a quantity that is considered highly informative, especially in the context of the pseudogap phase --, but hardly accessible in this regime using methods like MPS.  \\

\begin{figure}[t]
\centering
\includegraphics[width=0.49\textwidth]{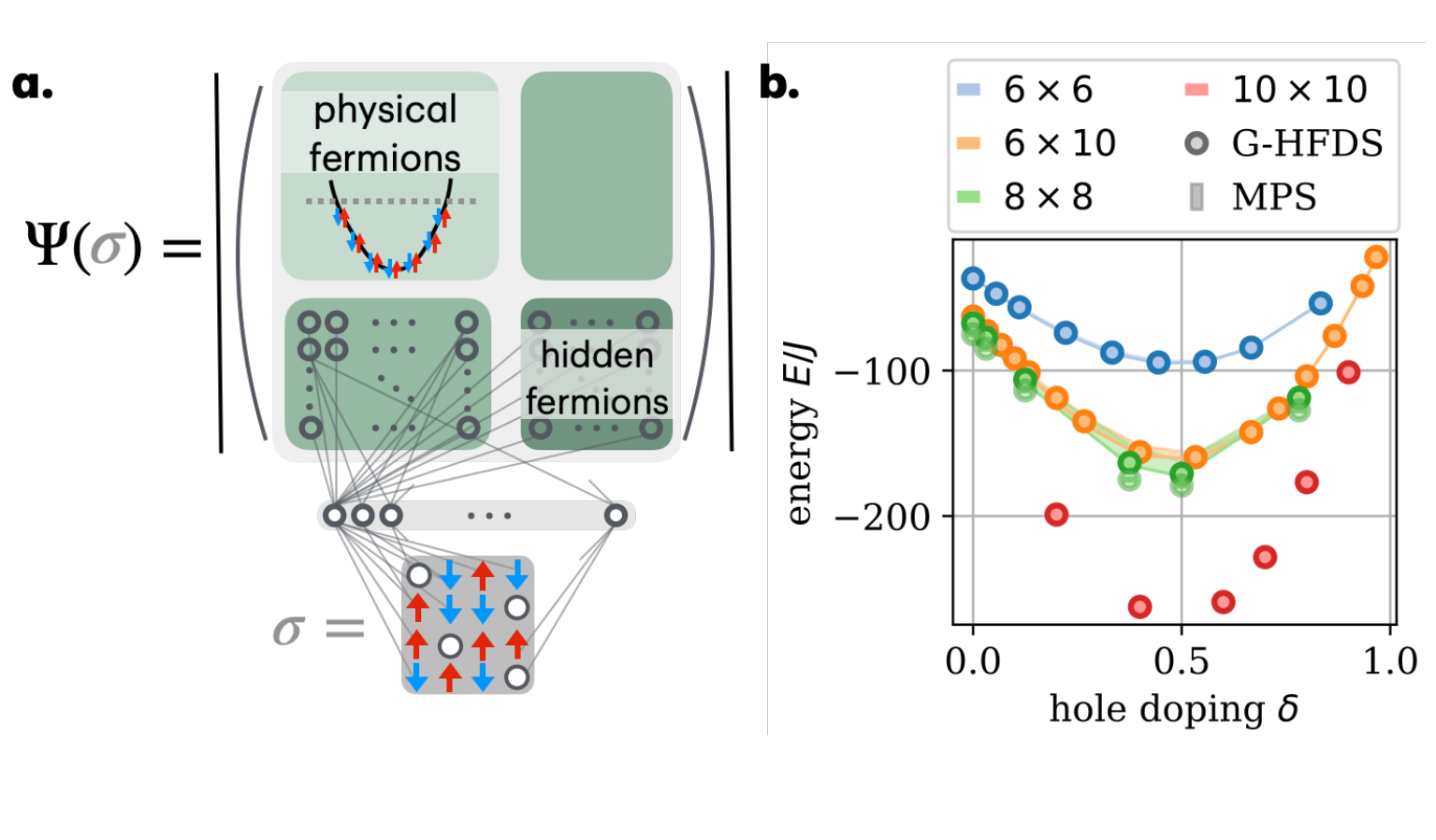}
\caption{\textbf{a.} Hidden fermion determinant states (HFDS): The wave function coefficients $\psi(\sigma)$ in the Fock basis $\sigma$ are parameterized as a determinant of physical and ancilla (\textit{hidden}) orbitals. The orbitals of the latter are configuration dependent, with the dependence learned by a feed-forward neural network. \textbf{b.} The energies for the $t$-$J$ model on open boundary $6\times 6$ (blue), $6\times 10$ (orange), $8\times 8$ (green) and $10\times 10$ (red) lattices at $t/J=3$ and different hole doping $\delta$, obtained from MPS (shaded regions for energies obtained with bond dimensions $\chi=128,\dots ,2048$) and the G-HFDS (markers). Light green markers represent $8\times 8$ energies for periodic boundaries in both directions from the G-HFDS. Errorbars denoting the error of the mean are smaller than the markers.}
\label{fig:1}
\end{figure}

\textit{Neural network architecture.--} We introduce G-HFDS, combining the HFDS \cite{Moreno2022},
and the concept of Gutzwiller projected wave functions \cite{GROS198953,Marston1989,DallaPiaza2015,DallaPiazaThesis}: For a sampled set of Fock space configurations $\{\vec{\sigma}\}$, we approximate 
\begin{align}
    \ket{\Psi} \approx \sum_{\{\vec{\sigma}\}} \psi_\mathrm{G-HFDS}(\sigma) \ket{\sigma}.
\end{align}
with $\psi_\mathrm{G-HFDS}(\vec{\sigma}) =\mathcal{P}_G \psi(\vec{\sigma})$. The former is based on a Slater determinant of both physical and auxiliary \textit{hidden} fermions, similar to Jastrow-factors \cite{Jastow1955} and neural backflow transformations \cite{Feynman1956,Luo2019,Hermann2020,pfau2020ferminet,kim2023neuralnetwork,romero2024spectroscopy}, see Supplementary Material (SM) \cite{SM} including Refs.~\cite{Slater1929,Stokes2020,Normura2017,Humeniuk2022,gauvinndiaye2023mott,liu2023unifying,wurst2024efficiencyhiddenfermiondeterminant,Barret2022,Inui2021,lange2023neural,Yoshioka2019}. The latter introduces a significant amount of correlations by restricting the sampling to configurations with at maximum singly occupied sites, effectively applying the projector $\hat{\mathcal{P}}_G$. Note that there is no additional computational cost in order to apply this projection, making G-HFDS an appealing candidate for simulating the $t$-$J$ model but also spin models, the Kugel-Khomskii or any model with a projection of this form. Furthermore, it allows to start the optimization from a pure Gutzwiller wave function by initializing the physical fermions accordingly.

The trainable upper matrix and the FFNN representing the lower matrix are then optimized using variational Monte Carlo (VMC) \cite{Becca2017,Goodfellow2016}, approximating an imaginary time evolution using the variant of stochastic reconfiguration introduced in Ref. \cite{Rende2024}. We additionally enforce a global spin flip symmetry.\\



For the HFDS, the augmentation of the physical Hilbert space leads to an enlarged Slater determinant $\psi(\vec{\sigma})$ consisting of four blocks, see Fig. \ref{fig:1}\textbf{a.}: The upper left (lower right) block represents the single particle orbitals of the physical (hidden) fermions. Respectively, the off-diagonal blocks represent interactions between the physical and auxiliary states. Hereby, the upper (lower) part of the matrix is parameterized by a set of trainable real-valued parameters (by a (configuration dependent) feed forward neural network) \cite{Moreno2022}. We start by initializing the physical block with free fermion orbitals and setting the hidden block (off-diagonal blocks) to identity (zero), resulting in the well-known variational ansatz for $t$-$J$ systems of Gutzwiller projected wave functions \cite{GROS198953,Marston1989,DallaPiaza2015,DallaPiazaThesis}. 
\\

\begin{figure}[t]
\centering
\vspace{-0.1cm}
\includegraphics[width=0.49\textwidth]{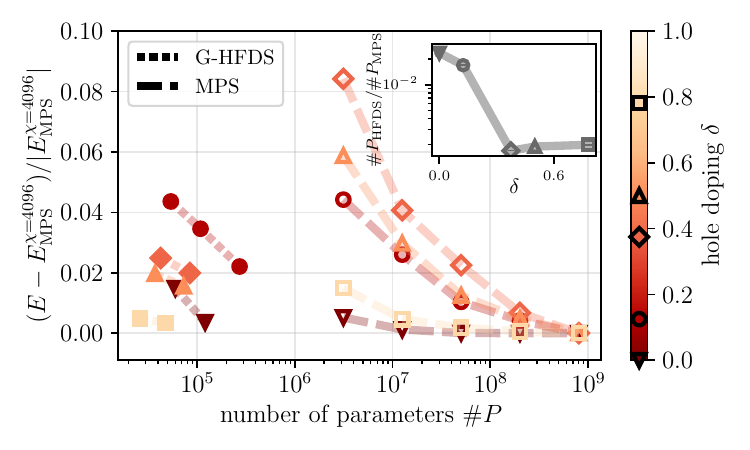}
\caption{Comparison of energies obtained from the G-HFDS (filled markers, dotted lines) and MPS (empty markers, dashed lines) for a $8\times 8$ lattice with open boundaries at exemplary dopings $\delta$ indicated by the colorbar. For the G-HFDS (MPS), the number of parameters \#$P_\mathrm{HFDS}$ (\#$P_\mathrm{MPS}$) is varied by the number of hidden fermions $N_h$ (bond dimension $\chi$). For the G-HFDS, errorbars denoting the error of the mean are smaller than the markers. For error estimates of the MPS results, see SM \cite{SM}. Inset: For each $\delta$, the number of parameters \#$P_\mathrm{HFDS}(\delta)\propto N_h$ for the highest considered $N_h$ is compared to the  \#$P_\mathrm{MPS}$ needed to obtain the same energy. The fraction \#$P_\mathrm{HFDS}/$\#$P_\mathrm{MPS}$ indicates the efficiency of the G-HFDS representation. }
\label{fig:MPSvsNQS}
\end{figure}

\textit{Comparison with MPS.--} The energies for the $t$-$J$ model are shown in Fig. \ref{fig:1}\textbf{b} for $8\times 8$ ($6\times 10$, $6\times 6$) lattices at different hole dopings $\delta$. All calculations are done for $t/J=3.0$ and $t^\prime=0$ unless stated differently. For comparison with MPS, open boundary systems are mainly considered. Note that, in contrast to MPS, HFDS can be used with both open, cylindrical and periodic boundary conditions in both directions without any difference in computational cost. As an example, we show the HFDS energies for a $8\times 8$ torus. We compare the open boundary results from the G-HFDS represented by the circles using a FFNN with a single hidden layer with $128$ ($128$, $78$) nodes and $20$ hidden fermions to $U(1)\times SU(2)$ symmetric matrix product calculations with bond dimensions $\chi=128,\dots,2048$, corresponding to up to $\approx 6000$ $U(1)$ states, which are represented by the shaded regions. In all cases, the results from G-HFDS and the MPS calculations with the highest bond dimension agree very well. We additionally calculate the ground state energies of a $10\times 10$ system with G-HFDS, which is extremely challenging for MPS methods.

Note that for intermediate dopings, the MPS energies for the large system strongly change between $\chi=128,\dots ,2048$ (shown by the shaded region), indicating that in this highly complicated regime with different competing energy scales $t$ and $J$ these bond dimensions are not sufficient to ensure convergence. In this regime, the power of G-HFDS states unfolds, as shown in more detail for $8\times 8$ systems at exemplary $\delta$ in Fig. \ref{fig:MPSvsNQS}: On the right side of Fig. \ref{fig:MPSvsNQS}, the energy over the number of parameters used by the MPS is shown, estimated as \#$P_\mathrm{MPS}=d\chi^2L_xL_y$, with $d=3$ the local dimension. At low and high doping, the MPS calculations converge for $\chi >512$ $SU(2)$ states, but for intermediate doping the energy significantly decreases up to the highest bond dimension $\chi=2048$ used here. We choose this as the maximal bond dimension in order to have a comparable number of G-HFDS parameters \#$P_\mathrm{HFDS}$ and MPS parameters \#$P_\mathrm{MPS}$. For the G-HFDS we choose to vary \#$P_\mathrm{HFDS}(\delta)\propto N_h^2$ at a fixed doping $\delta$ by changing the number of hidden fermions $N_h$. In all cases, the energies obtained with the G-HFDS (see Fig. \ref{fig:MPSvsNQS} left) agree well with the MPS energies, but require more than two orders of magnitude less parameters.\\ 

Focusing on the highest considered \#$P_\mathrm{HFDS}$ for each $\delta$, we calculate the \#$P_\mathrm{MPS}$ that is needed to obtain the same energy by fitting the MPS energies to a logarithmic dependence and determining the intersection of the fit with a horizontal line through the G-HFDS energies. The fraction \#$P_\mathrm{HFDS}/$\#$P_\mathrm{MPS}$ is shown in the inset. Again, it can be seen that the G-HFDS are particularly efficient in the intermediate doping regime. Furthermore, the training of the G-HFDS on GPUs with up to $1500$ steps took $6$ to $15$ GPU hours, while the MPS calculations ran several days on a CPU.   \\

\begin{figure}[t]
\centering
\includegraphics[width=0.49\textwidth]{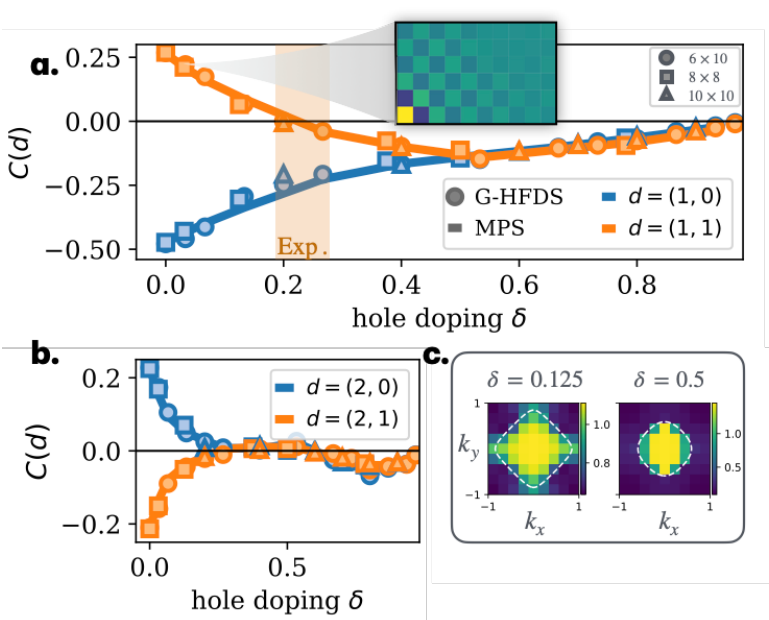}
\caption{\textbf{a.,b.} Two-point spin-spin correlators $C(\vec{d})$ for different distances $\vec{d}$ in the $6\times 10$ (circles), $8\times 8$ (squares), $10\times 10$ (triangles) lattice with open boundaries from the optimized G-HFDS. Lines represent results from MPS on $6\times 10$ systems. The sign change region from the experiment Ref.~\cite{Koepsell2021} is denoted by the orange region for $d=(1,1)$. Inset: Full spin correlation map for two holes and a reference spin in the lower left corner, colorbar limits are $-0.25$ (blue) to $0.25$ (yellow). Errorbars denoting the error of the mean are smaller than the markers. \textbf{c.} $k$-space densities $n_\mathbf{k}=n_{\mathbf{k}\uparrow}+n_{\mathbf{k}\downarrow}$ for exemplary dopings before and after the sign crossing. White dashed lines indicate the analytical result for the Fermi surface of free spin-$1/2$ fermions.}
\label{fig:spincorrs}
\end{figure}

\begin{figure*}[t]
\centering
\includegraphics[width=0.95\textwidth]{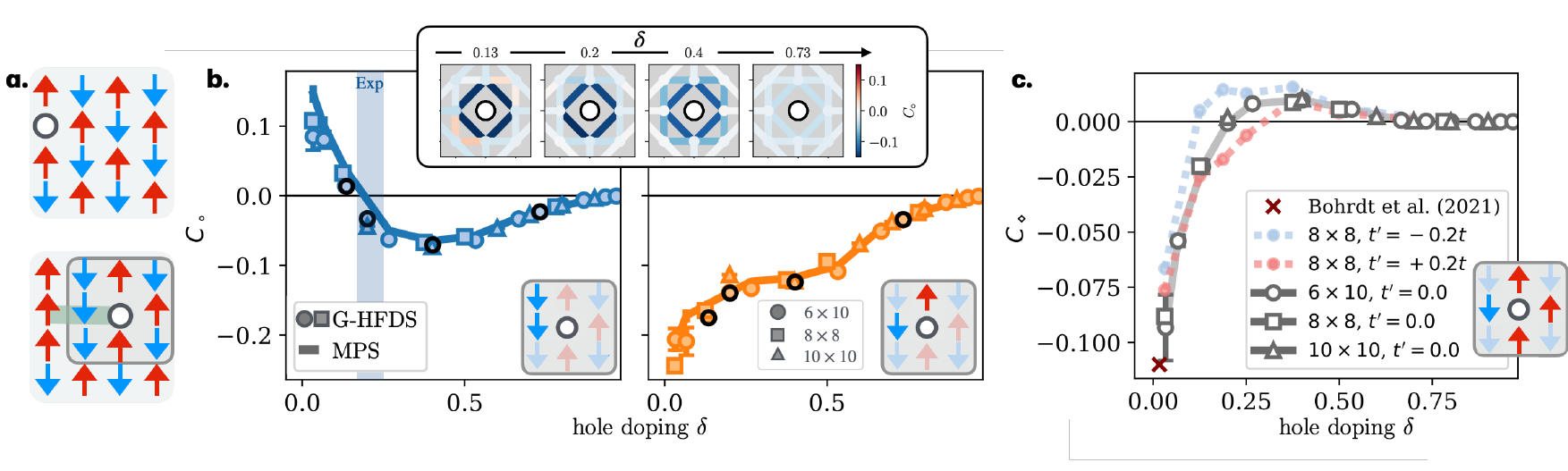}
\caption{Investigating the polaron regime. \textbf{a.} Geometric string theory: When the hole (represented by the circle) moves away from its original position (top), it distorts the magnetic background (blue and red arrows), giving rise to an energy cost that increases linearly with the distance; a \textit{string} (green) of length $l_\Sigma$. \textbf{b}. Three-point polaron correlators $C_\circ(\vec{d},\vec{r}_h)$ for $6\times 10$ (circles), $8\times 8$ (squares) and $10\times 10$ (triangles) systems with open boundaries and different distances $\vec{d}=(0,1),(1,1)$ (left, right). Circles represent data obtained from the optimized NQS wavefunction, lines $6\times 10$ MPS calculations. The sign change region from the experiment Ref.~\cite{Koepsell2021} is denoted by the blue region for $d=(1,0)$. Black markers indicate the doping values for which the full correlation map for the $6\times 10$ lattice is shown in the inset.
\textbf{c}. Five-point polaron correlations $C_\diamond$ for $6\times 10$, $8\times 8$ and $10\times 10$ systems with open boundaries for different next-nearest neighbor hoppings $t^\prime$. For a single hole, we compare to the result from Ref. \cite{Bohrdt2020}. Errorbars (mostly smaller than the markers) show the error of the mean. The position of the hole relative to the spins $\vec{r}_h$ is shown in the insets for \textbf{b} and \textbf{c}. In the very low doping regime, positions with $\langle \hat{n}_\vec{r}\rangle =0$ exist, leading to effectively larger errors of both $C_\circ$ and $C_\diamond$. These errors are estimated by two different realizations of the G-HFDS, see main text.}
\label{fig:polaroncorrs}
\end{figure*}

\textit{Full doping scans.--} In order to investigate the evolution of the interplay between magnetic and kinetic degrees of freedom with doping, e.g. the extent of the polaron regime, full doping scans of low energy state physical observables like correlation functions are needed. However, while at finite temperature, spin and polaron correlation functions have been measured e.g. on quantum simulation platforms realizing the Hubbard model \cite{Koepsell2021,Cheuk2016}, accessing low energy or ground state observables in the full doping regime  has remained experimentally challenging and computationally expensive, particularly at finite dopings. With the efficiency of our G-HFDS ansatz in the intermediate doping regime, such doping scans become feasible. In the following, we compare our low energy two-point spin and three-point polaron correlations to experimental measurements at finite temperature and make predictions for experimentally feasible five-point polaron correlations. In all cases, we take averages over the full system, excluding terms at the boundaries where the expectation values would involve sites outside the samples. \\

We start by analyzing the connected spin-spin correlation function between spins at sites $\boldsymbol{r}_1$ and $\boldsymbol{r}_2$,
\begin{align}
    C(\vec{d}) = \frac{1}{N_d} \sum_{\substack{\vec{r}_1,\vec{r}_2\,\mathrm{s.t.}\\\vec{r}_1-\vec{r}_2=\vec{d}}} \eta_{\vec{r}_1\vec{r}_2} \langle \hat{S}^z_{\vec{r}_1}\hat{S}^z_{\vec{r}_2}\rangle_c,
    \label{eq:spinspincorr}
\end{align}
where $N_d$ is the number of terms in the sum, $\eta_{\vec{r}_1\vec{r}_2}=\left[\sigma(\hat{S}^z_{\vec{r}_1}) \sigma(\hat{S}^z_{\vec{r}_2}) \right]^{-1}$ a normalization accounting for the increased presence of holes at higher doping levels (with the standard deviation $\sigma$) and $\langle \dots \rangle_c$ denotes the connected correlator with lower-order terms subtracted, as defined in SM \cite{SM}. The results for different values of $d$ are displayed in Fig.~\ref{fig:spincorrs}, and compared to MPS results ($T=0$) and experimental measurements from Ref. \cite{Koepsell2021}, taken at temperature $k_BT=0.4t$ in the Fermi-Hubbard model. Similar measurements were also taken in Ref. \cite{Cheuk2016}. The results for MPS and G-HFDS agree very well: At half-filling, $d=(1,0)$ reveals a clear antiferromagnetic alignment of spins, while we observe ferromagnetic correlations at $d=(1,1)$. Upon doping, both correlations become weaker. We see the same behavior for spins at larger distances $d=(2,0), (2,1)$, consistent with AFM N\'eel order at half-filling, which quickly decreases with higher doping.

At doping of $\delta\approx0.25$, some correlations undergo a sign change. This sign change is associated with a breakdown of the magnetic polaron regime and a crossover to a Fermi liquid and also observed in the experiment (shaded region) \cite{Koepsell2021}. The crossover is further associated with a qualitative change in the Fermi surface (FS): In the low-doping regime, the charge carriers are free magnetic polarons, associated with a small FS and hole pockets, while the Fermi liquid has a large FS of free fermions. We analyze the Fermi surface from the occupation in $k$-space given as $\langle n_{\mathbf{k}\sigma}\rangle=\frac{1}{L^2}\sum_{j,j'}e^{i\mathbf{k}(j-j')}\langle c_{j\sigma}^\dagger c_{j'\sigma}\rangle$, results are shown in Fig.~\ref{fig:spincorrs}c. For dopings above the sign change $\delta>0.25$, we find a FS closely aligned with free fermions, while for $\delta<0.25$ before the sign change, we find a qualitative change, where the distribution is shifted towards $(\pm\pi,0)$ and $(0,\pm\pi)$ compared to the free fermion FS.
The sign change can also be understood in terms of the geometric string theory~\cite{Beran1996,Grusdt2018, Bohrdt2021}, where motion of a single dopant in a magnetic background creates a string $\Sigma$ of displaced spins, see Fig. \ref{fig:polaroncorrs}\textbf{a}.  Beyond a single dopant, this description is still valid in the low doping regime~\cite{Chiu2019}, but with an increasing number of dopants, strings start to mix -- and hence also the corresponding correlations. Once a certain level of mixing is reached, this introduces a sign change in some correlations. Note that the full correlation map does not show any signature of a stripe phase in the low doping regime, see inset of Fig. \ref{fig:spincorrs}, as e.g. discussed in Refs. \cite{White1997,White1999,Jiang2021}. This is different for cylindrical boundaries, see SM \cite{SM}.\\

The extent of the magnetic polaron regime can be studied in more detail by considering correlations of two spins around a hole, see also Ref. \cite{Bohrdt2021}, 
\begin{align}
    C_\circ(\vec{d},\vec{r}_h) =  \frac{1}{N_d N_{r_h}}\sum_{\vec{r}_h}\sum_{\substack{\vec{r}_1,\vec{r}_2\, \mathrm{s.t.} \\\,\vec{r}_1-\vec{r}_2=\vec{d}\\\,\vert \vec{r}_1-\vec{r}_h\vert =1}} \Tilde{\eta} \langle \hat{S}^z_{\vec{r}_1}\hat{S}^z_{\vec{r}_2}\hat{n}^h_{\vec{r}_h}\rangle_c,
    \label{eq:3pointpolaroncorr}
\end{align}
with the hole density $\hat{n}^h_{\vec{r}_h}$  and $\Tilde{\eta}=\left[\langle \hat{n}^h_{\vec{r}_h} \rangle\sigma(\hat{S}^z_{\vec{r}_1}) \sigma(\hat{S}^z_{\vec{r}_2}) \right]^{-1}$, measuring the influence of the hole on the spin correlations in its vicinity. The results from G-HFDS, MPS and experiment \cite{Koepsell2021} are shown in  Fig. \ref{fig:polaroncorrs}\textbf{b}.
For low doping, the antiferromagnetic background is perturbed by the hole as schematically illustrated in Fig. \ref{fig:polaroncorrs}\textbf{a}, i.e. nearest neighbor ($d=(1,0)$) spins align more ferromagnetically, and diagonal neighbors ($d=(1,1)$) more antiferromagnetically. At low doping, the absolute strength of magnetic correlations decreases due to the presence of holes as observed before. Note that \eqref{eq:3pointpolaroncorr} is not well defined if $\langle \hat{n}^h_{\vec{r}_h}\rangle=0 $. In these cases, we set the contribution of the respective term to the sum to zero. For the very low hole dopings where this problem occurs for the $6\times 10$ systems (marked by the gray area), we encounter higher errorbars, estimated by the results from two different parameterizations of the state. All other error bars indicate the sampling error.

Furthermore, again a sign change in the correlations with $d=(1,0)$ occurs, indicating a transition from the strongly correlated state described by magnetic string theory, to a Fermi Liquid, where the Pauli exclusion principle leads to enhanced antiferromagnetic alignment~\cite{Hartke_2020}. For cuprate materials, the same qualitative change of observables takes place at a similar doping level \cite{Keimer2015}. Note that the presence of real space pairs would lead to similar signatures as observed here for the polaron correlations \cite{White1997}.\\

To further investigate the nature of the polaron, the five-point correlator
\begin{align}
    C_\diamond = \frac{2^4}{N_r}\sum_{\vec{r}_h}\frac{1}{\langle \hat{n}^h_{\vec{r_h}}\rangle}\langle \hat{n}^h_{\vec{r_h}}\hat{S}^z_{\vec{r_h}+\hat{e}_x}\hat{S}^z_{\vec{r_h}-\hat{e}_x}\hat{S}^z_{\vec{r_h}+\hat{e}_y}\hat{S}^z_{\vec{r_h}-\hat{e}_y}\rangle
    \label{eq:5pointpolaroncorr}
\end{align}
is considered \cite{Bohrdt2020} in Fig. \ref{fig:polaroncorrs}\textbf{c}. At very low doping, $C_\diamond$ depends on the probability of configurations with string lengths $l_\Sigma= 0$ and $l_\Sigma> 0$. As schematically shown in Fig. \ref{fig:polaroncorrs}\textbf{a}, $C_\diamond>0$ for $l_\Sigma= 0$, while for $l_\Sigma>0$, $C_\diamond<0$. The fact that at low doping we find $C_\diamond<0$ indicates that configurations with $l_\Sigma>0$ are more likely, in agreement with Ref. \cite{Grusdt2019}. Upon increasing the doping, $C_\diamond$ become positive with the sign change depending on the next-nearest neighbor hopping amplitude $t^\prime$. This is in agreement with the faster decay of AFM correlations w.r.t. doping for hole ($t^\prime <0$) vs. particle ($t^\prime>0$) doping \cite{Keimer2015}.\\

\textit{Conclusion.--} In this work, we demonstrate that Gutzwiller projected hidden fermion determinant states (G-HFDS) effectively capture the low-energy states of the $t$-$J$ model. Particularly, we show that G-HFDS can encode energies comparable to those obtained from matrix product state (MPS) calculations with $2048$ $U(1)\times SU(2)$ symmetric states across the full doping regime, while utilizing up to four orders of magnitude fewer parameters. Furthermore, they can be used with periodic boundary conditions. This parameter efficiency positions G-HFDS as a competitive alternative to MPS, particularly as GPU resources continue to advance.

In addition to improving training speed, reducing parameter count makes it easier to save model weights as they take up far less memory. As a result, we are able to share model weights across the full doping scan so that others can access the trained models. The model weights can either be used to compute new correlation functions, or as a starting point for simulating closely related models.  

Our analysis reveals that G-HFDS excel in the challenging intermediate doping regime, facilitating the exploration of low-energy observables throughout the entire doping scan. Specifically, we investigate spin and polaron correlations, comparing our findings with experimental data and MPS calculations. The doping dependence of these correlation functions, and the critical dopings at which their signs change, aligns well with existing literature. This work paves the way for further investigations of correlated systems across the full doping range and low temperatures. \\

\textbf{Code availability.}  Our implementation of hidden fermions is adapted from the neural backflow \cite{NetKetLatticeFermions} and is available at \cite{HFDSfortJ}, where also the trained models can be found.\\

\textbf{Acknowledgements.} We thank Fabian Grusdt, Tizian Blatz, Timothy Harris, and Subir Sachdev for useful discussions.  All DMRG calculations were performed using the SyTeN toolkit \cite{syten1,syten2} developed and maintained by C. Hubig, F. Lachenmaier, N.-O. Linden, T. Reinhard, L. Stenzel, A. Swoboda, M. Grundner, S. Mardazad, F. Pauw, and S. Paeckel. Information is available at \url{https://syten.eu} and in SM \cite{SM}. A. Bohrdt, A. Böhler and H. Lange acknowledge the support by the Deutsche Forschungsgemeinschaft (DFG, German Research Foundation) under Germany’s Excellence Strategy—EXC-2111—390814868. H. Lange acknowledges support by the International Max Planck Research School for Quantum Science and Technology (IMPRS-QST). \\C. Roth acknowledges support from the Flatiron Institute. The Flatiron Institute is a division of the Simons Foundation.

\bibliography{main}
\clearpage

\newpage
\appendix
\onecolumngrid
\begin{center}
\textbf{\Large Supplementary Materials}
\end{center}


\setcounter{equation}{0}
\setcounter{figure}{0}
\setcounter{table}{0}
\setcounter{page}{1}
\makeatletter
\renewcommand{\theequation}{S\arabic{equation}}
\renewcommand{\thefigure}{S\arabic{figure}}

\section{The Hidden Fermion Determinant States}

\subsection{Simulating fermionic systems with neural quantum states}

While for non-interacting fermions, the wave function can be constructed as the Slater determinant of single-particle orbitals \cite{Slater1929}, for interacting fermions, the wave function coefficients depend on the position of all particles. To take this configuration dependence into account, the non-interacting Slater determinant can be dressed in various ways. Hereby, the capability of neural networks to represent wide classes of functions  has proven to be useful to learn the configuration dependence by neural networks \cite{medvidović2024neuralnetwork,lange2024architecturesapplicationsreviewneural} on top of the fermionic sign structure provided by the Slater determinant. Currently, there are three approaches to incorporate the fermionic sign structure into the nerual network ansatz: $(i)$ Jastrow-factors \cite{Jastow1955}, i.e. the Slater determinant is multiplied with a neural network correlation factor \cite{Nomura2017,Stokes2020,Humeniuk2022,nomura2024quantummanybodysolverusing}. $(ii)$ Neural backflow transformations, dressing the single-particle orbitals by a (configuration-dependent) network  \cite{Feynman1956,Luo2019,Hermann2020,pfau2020ferminet,kim2023neuralnetwork,romero2024spectroscopy}. $(iii)$ Hidden fermion determinant states (HFDS), where the Slater determinant is enlarged and includes the physical single particle orbitals as well as single particle orbitals from additional projected \textit{hidden} fermions, with their \textit{hidden} orbitals represented by neural networks \cite{Moreno2022,gauvinndiaye2023mott}. Both neural backflow and HFDS can be rewritten as a Jastrow-like corrections to the single particle Slater determinant \cite{liu2023unifying} and their efficiency in representing volume-law entangled states has been investigated in Ref. \cite{wurst2024efficiencyhiddenfermiondeterminant}. Lastly, we would like to mention that fermions can also be simulated using fermionic mappings in combination with NQS, see e.g. \cite{lange2024architecturesapplicationsreviewneural}.\\

The G-HFDS that we introduce in this work combine the HFDS \cite{Moreno2022},
and the concept of Gutzwiller projected wave functions \cite{GROS198953,Marston1989,DallaPiaza2015,DallaPiazaThesis}. In contrast to neural quantum states based on fermionic mappings \cite{Barret2022,Yoshioka2019,Inui2021,lange2023neural}, this representation inherently possesses the fermionic sign structure. 
We additionally enforce a global spin flip symmetry by defining $\psi(\sigma)_\mathrm{sym}= s(\sigma)\frac{1}{2}(\Tilde{\psi}(\sigma) +\Tilde{\psi}(\mathcal{T}\sigma))$, where the real-valued determinant $\psi(\sigma)=s(\sigma)\Tilde{\psi}(\sigma)$ is split into its sign $s(\sigma)$ and $\Tilde{\psi}(\sigma)$, and $\mathcal{T}$ flips all spins of $\sigma$.\\

While the HFDS can in principle represent systems where double occupancy of spin up and down fermions is allowed, we apply the projector $\hat{\mathcal{P}}_G$ by initializing the Monte Carlo chain without doubly occupied configurations and avoid the appearance later in the chain by appropriate updates. Even for a single Slater determinant without any trainable parameters, constructed of free fermion orbitals, this procedure introduces a significant amount of correlations, and are a well-known variational ansatz for $t$-$J$ systems under the name of Gutzwiller projected wave functions \cite{GROS198953,Marston1989,DallaPiaza2015,DallaPiazaThesis}. Furthermore, this physical knowledge can accelerate the convergence and reduce the computational effort.

\section{Definitions and Additional Results}

\subsection{Convergence for the $6\times 10$ systems}
Below, we show the same convergence analysis as in Fig. 2 for the $6 \times 10$ systems. Both the $6 \times 10$ as well as the $8 \times 8$ systems were trained on 1-2 $H100$ GPUs.

\begin{figure}[htp]
\centering
\includegraphics[width=0.49\textwidth]{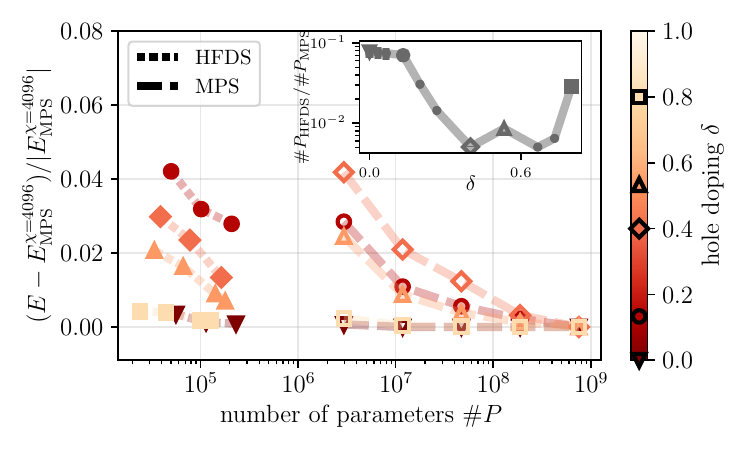}
\caption{Comparison of energies obtained from the G-HFDS (filled markers, dotted lines) and MPS (empty markers, dashed lines) for a $6\times 10$ lattice at exemplary dopings $\delta$ indicated by the colorbar. For the G-HFDS (MPS), the number of parameters \#$P_\mathrm{HFDS}$ (\#$P_\mathrm{MPS}$) is varied by the number of hidden fermions $N_h$ (bond dimension $\chi$). For the G-HFDS, errorbars denoting the error of the mean are smaller than the markers. Inset: For each $\delta$, the number of parameters \#$P_\mathrm{HFDS}(\delta)\propto N_h$ for the highest considered $N_h$ is compared to the  \#$P_\mathrm{MPS}$ needed to obtain the same energy. The fraction \#$P_\mathrm{HFDS}/$\#$P_\mathrm{MPS}$ indicates the efficiency of the G-HFDS representation. The G-HFDS results are obtained without enforcing the spin flip symmetry discussed in the main text.}
\label{fig:MPSvsNQS_6x10}
\end{figure}

\subsection{Definition of the connected correlation functions \label{sec:Corrs}}
The two-point spin-spin correlations (2), as well as the three-point polaron correlations (3) shown in the main text are defined as the connected versions. When writing out the connected correlators $\langle \dots \rangle_c$ used in the main text, they take the following forms:
\begin{itemize}
    \item The two-point spin-spin correlations are
    \begin{align}
    C(\vec{d}) = \frac{1}{N_d} \sum_{\vec{r}_1,\vec{r}_2\,\mathrm{s.t.}\,\vec{r}_1-\vec{r}_2=\vec{d}} \eta_{\vec{r}_1\vec{r}_2}\left( \langle \hat{S}^z_{\vec{r}_1}\hat{S}^z_{\vec{r}_2}\rangle-\langle \hat{S}^z_{\vec{r}_1}\rangle\langle \hat{S}^z_{\vec{r}_2}\rangle \right)
    \end{align}
    \item The three-point polaron correlations are
    \begin{align}
    C_\circ(\vec{d},\vec{r}_h) =  \frac{1}{N_d N_{r_h}}\sum_{\vec{r}_h}\sum_{\vec{r}_1,\vec{r}_2\,\mathrm{s.t.}\,\vec{r}_1-\vec{r}_2=\vec{d}} \Tilde{\eta}_{\vec{r}_1\vec{r}_2\vec{r}_h} \Big(& \langle \hat{S}^z_{\vec{r}_1}\hat{S}^z_{\vec{r}_2}\hat{n}_{\vec{r}_h}\rangle - \langle \hat{S}^z_{\vec{r}_1}\rangle \langle \hat{S}^z_{\vec{r}_2}\hat{n}_{\vec{r}_h}\rangle -\langle \hat{S}^z_{\vec{r}_1}\hat{S}^z_{\vec{r}_2}\rangle \langle\hat{n}_{\vec{r}_h}\rangle\notag \\ &-\langle\hat{S}^z_{\vec{r}_2} \rangle \langle\hat{S}^z_{\vec{r}_1}\hat{n}_{\vec{r}_h}\rangle+2\langle \hat{S}^z_{\vec{r}_1}\rangle \langle \hat{S}^z_{\vec{r}_2}\rangle \langle\hat{n}_{\vec{r}_h}\rangle\Big)
\end{align}
\end{itemize}

We show additional results for these correlators below:
\subsubsection{Additional correlations}
\begin{figure}[t]
\centering
\includegraphics[width=0.9\textwidth]{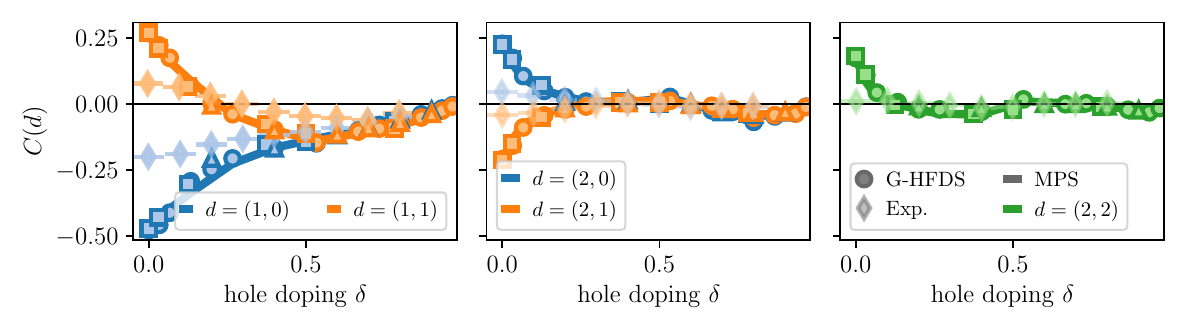}
\includegraphics[width=0.9\textwidth]{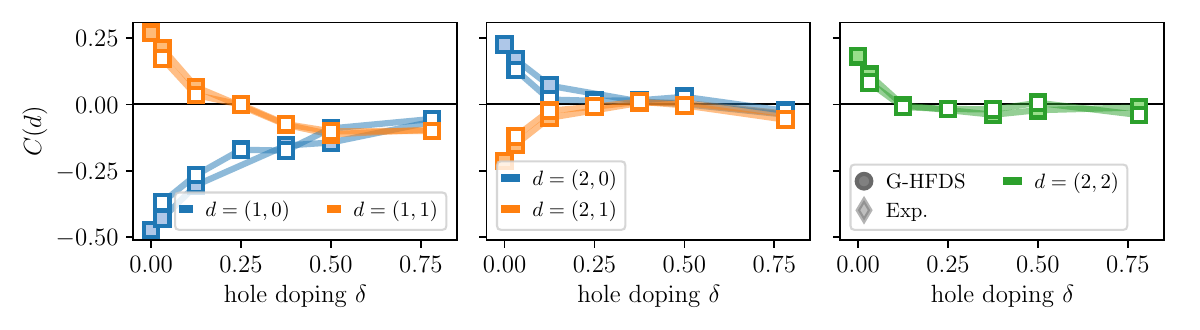}
\caption{Spin-spin correlations $C(\vec{d})$ for different distances  $\vec{d}=(1,0),(1,1)$ (left), $\vec{d}=(2,0),(2,1)$ (middle) and $\vec{d}=(2,2)$ (right). \textbf{Top: Comparison for different system sizes with Experiment} For G-HFDS, we consider systems of size $6\times 10$ (circles), $8\times 8$ (squares) and $10\times 10$ (triangles) and compare to MPS of size $6\times 10$ (lines). We furthermore compare to experimental data from Ref. \cite{Koepsell2021} which is taken at finite temperature and for the Fermi Hubbard model. \textbf{Bottom: Comparison for different boundary conditions} G-HFDS comparison for $8\times 8$ systems with cylindrical (blue markers) and fully periodic boundaries (white markers).}
\label{fig:spincorrs_add}
\end{figure}

The results for the spin correlations $C(\vec{d})$ for different distances  $\vec{d}=(1,0),(1,1)$, $\vec{d}=(2,0),(2,1)$ and $\vec{d}=(2,2)$ (right) are shown in Fig. \ref{fig:spincorrs_add} for MPS, G-HFDS and the experiment. The top panel shows the comparison for different system sizes from G-HFDS ($6\times 10$ (circles), $8\times 8$ (squares) and $10\times 10$ (triangles)) and MPS ($6\times 10$ (lines)) with the experimental data from Ref. \cite{Koepsell2021}. Note that this data is taken at finite temperature and for the full Fermi Hubbard model and hence we dont expect quantitative agreement. However, the sign change, indicating a qualitative change of the physics, agrees with our numerics. 

The bottom panel shows a comparison pf G-HFDS for $8\times 8$ systems with cylindrical (blue markers) and fully periodic boundaries (white markers), where we find good agreement.

\begin{figure}[t]
\centering
\includegraphics[width=0.85\textwidth]{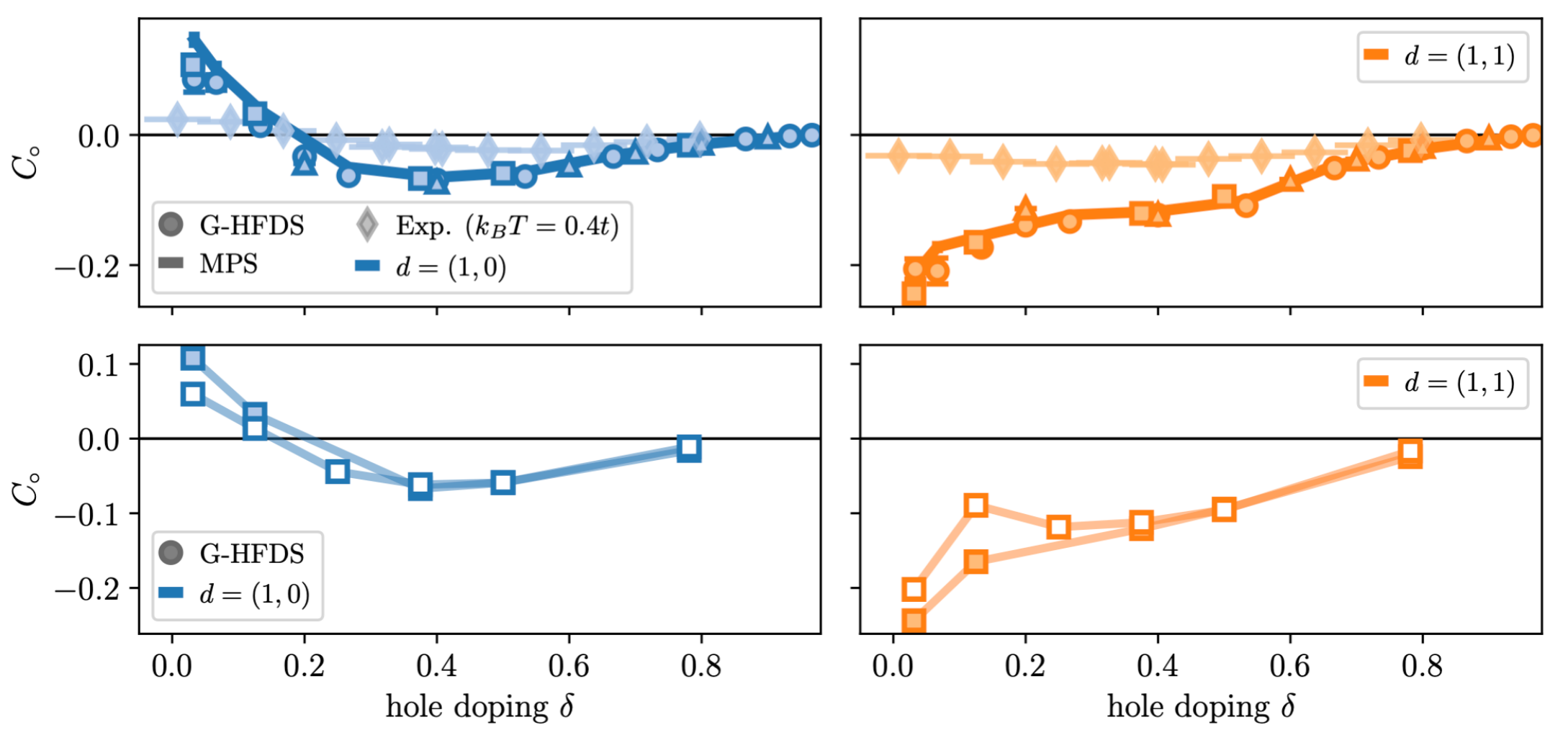}
\caption{Polaron correlations $C(\vec{d})$ for different distances  $\vec{d}=(1,0),(1,1)$. \textbf{Top: Comparison for different system sizes with Experiment} For G-HFDS, we consider systems of size $6\times 10$ (circles), $8\times 8$ (squares) and $10\times 10$ (triangles) and compare to MPS of size $6\times 10$ (lines). We furthermore compare to experimental data from Ref. \cite{Koepsell2021} which is taken at finite temperature and for the Fermi Hubbard model. \textbf{Bottom: Comparison for different boundary conditions} G-HFDS comparison for $8\times 8$ systems with cylindrical (blue markers) and fully periodic boundaries (white markers).}
\label{fig:polaroncorrs_add}
\end{figure}

The results for the polaron correlations are shown in Fig. \ref{fig:polaroncorrs_add}. The results are very similar as for the spin correlations. For low doping and $d=(1,1)$ the correlations slightly deviate for the two considered boundary conditions, which could be resulting from the convergence of the fully periodic systems. 
\FloatBarrier

\subsection{Discussion of the stripe phase \label{sec:Stripes}}

\begin{figure}[b]
\centering
\includegraphics[width=0.49\textwidth]{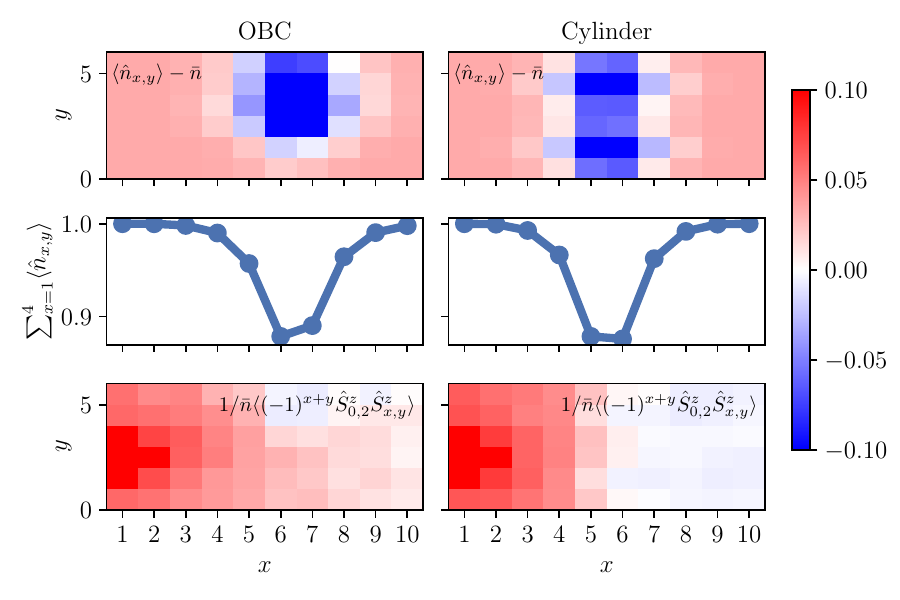}
\includegraphics[width=0.49\textwidth]{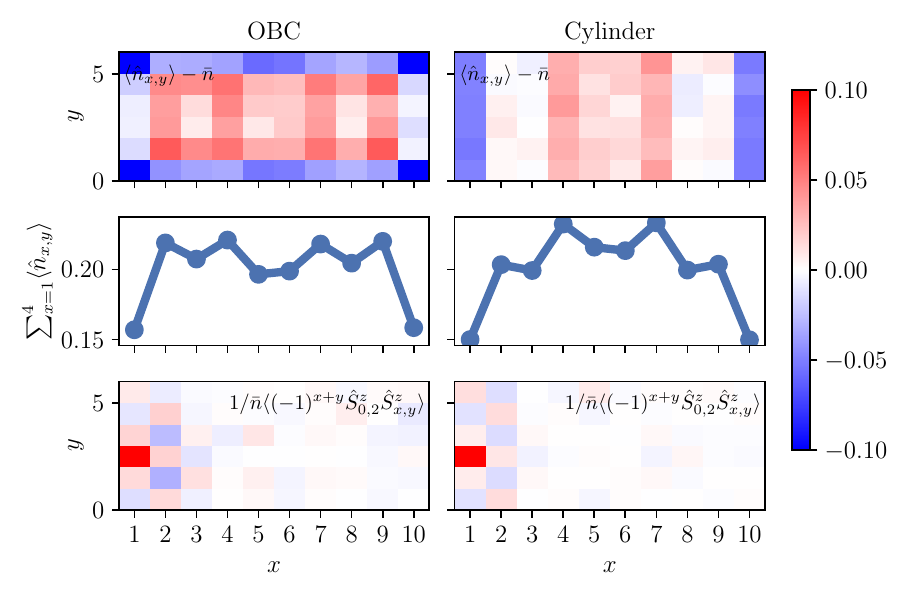}
\caption{Open and cylindrical boundaries: We show the local density $\langle \hat{n}_{x,y}\rangle -\Bar{n}$ with $\Bar{n}=1-\delta$ (top), the average rung density (middle) and the staggered and normalized spin correlations $1/\bar{n}  \langle (-1)^{x+y} \hat{S}^z_{0,2}\hat{S}^z_{x,y}\rangle $ for systems with hole doping $\delta=0.03$ ($N=58$, left) and $\delta=0.80$ ($N=12$, right), comparing open (left column) and cylindrical (i.e. open in the long direction, periodic in the short direction, right column) boundaries.  }
\label{fig:density}
\end{figure}

Below, we show that while there are no clear signatures for stripes for the open boundary systems considered in the main text, cylindrical boundaries (i.e. open in the long $x$ direction, periodic in the short $y$ direction) can give rise to stripe signals in the density and spin correlations.

In Fig. \ref{fig:density} we show the local density $\langle \hat{n}_{x,y}\rangle -\Bar{n}$, subtracted by the average filling $\Bar{n}=1-\delta$ as well as the average rung density for both boundary conditions and exemplary hole doping $\delta=0.03$ ($N=58$, left) and $\delta=0.80$ ($N=12$, right) in the top two rows. For both boundary conditions, we find a modulation of the density in the $y$ direction, with slightly more pronounced peaks for the cylindrical boundaries at $\delta=0.80$. Furthermore, we calculate the staggered and normalized spin correlations $$\frac{1}{\bar{n}}  \langle (-1)^{x+y} \hat{S}^z_{0,2}\hat{S}^z_{x,y}\rangle. $$ 
For the cylindrical boundaries, we find that the signal of these correlations is suppressed when $x,y$ are behind a density modulation with filling lower than the average, most prominently for $\delta=0.03$, indicating a stripe-like phase.

\subsection{Hole-hole correlations}
\begin{figure}[t]
\centering
\includegraphics[width=0.9\textwidth]{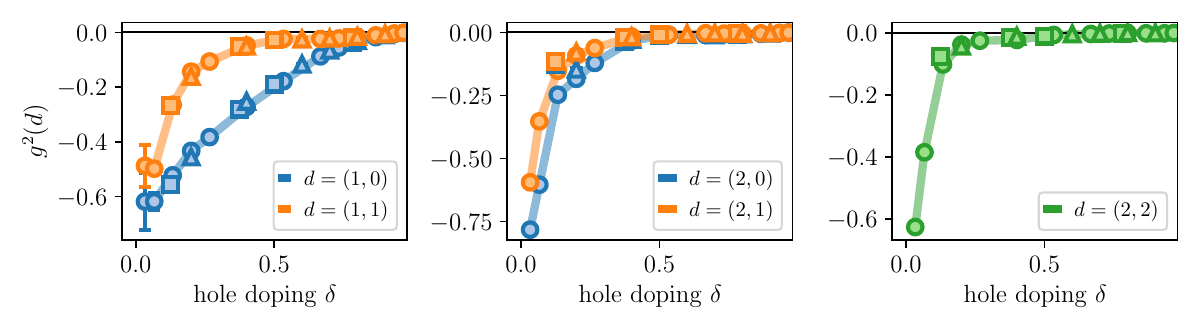}
\caption{Hole-hole correlations $g^2(\vec{d})$ as defined in Eq. \eqref{eq:holecorrs} for different distances  $\vec{d}=(1,0),(1,1)$ (left), $\vec{d}=(2,0),(2,1)$ (middle) and $\vec{d}=(2,2)$ (right) and systems of size $6\times 10$ (circles), $8\times 8$ (squares) and $10\times 10$ (triangles). }
\label{fig:holecorrs}
\end{figure}

Fig. \ref{fig:holecorrs} shows the hole-hole correlations
\begin{align}
    g^2(\vec{d})=\sum_{\vec{r}_i, \,\vec{r}_j,\mathrm{s.t.}\,\vec{r}_i, \vec{r}_j=\vec{d}}\left(\langle\hat{n}_{\vec{r}_i}  \hat{n}_{\vec{r}_j} \rangle/ (\langle \hat{n}_{\vec{r}_i} \rangle  \langle \hat{n}_{\vec{r}_j} \rangle) \right)-1
    \label{eq:holecorrs}
\end{align}
across the full doping range. In all cases, the observed hole-hole correlations are negative i.e. we do not observe any signatures of hole bunching, which in principle could arise from
interactions between the holes mediated by the spin
background. This is in agreement with the experimental data from Ref. \cite{Koepsell2021}.

\subsection{Momentum space distributions}

\begin{figure}
    \centering
    \includegraphics[width=0.99\linewidth]{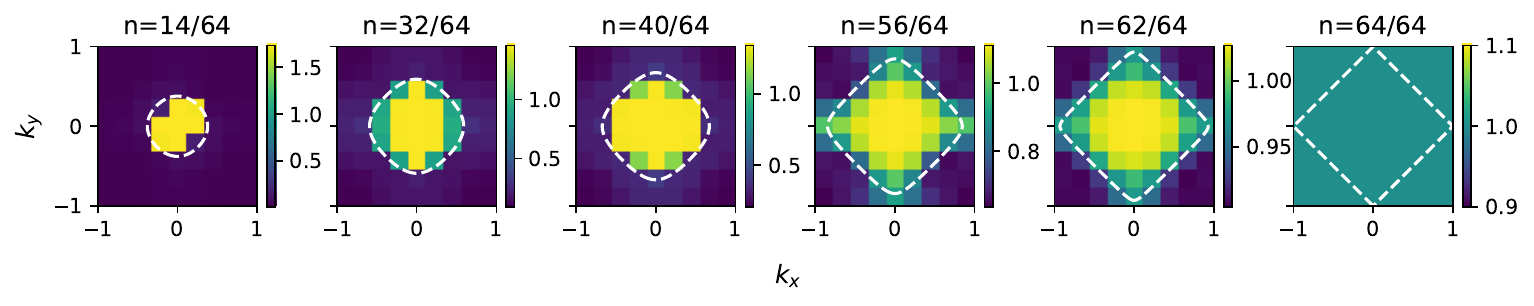}
    \caption{Momentum space distribution $n_k$ for different dopings $\delta$. Dashed white lines represent the exact expression for free fermions.}
    \label{fig:FermiSurface}
\end{figure}

Fig~\ref{fig:FermiSurface} shows the momentum space distribution $n_k=\langle \hat{c}_k^\dagger\hat{c}_k\rangle$ for a full doping scan. We compare to the analytical expectation of the Fermi surface of free fermions, analogous to $\delta=0.125$ and $\delta=0.5$ shown also in the main text. We observe a good agreement with the free fermion Fermi surface at high dopings, while deviations occur at low dopings, where the system is far from a Fermi liquid and in the magnetic polaron regime.

\subsection{Density scans}

\begin{figure}
    \centering
    \includegraphics[width=\textwidth]{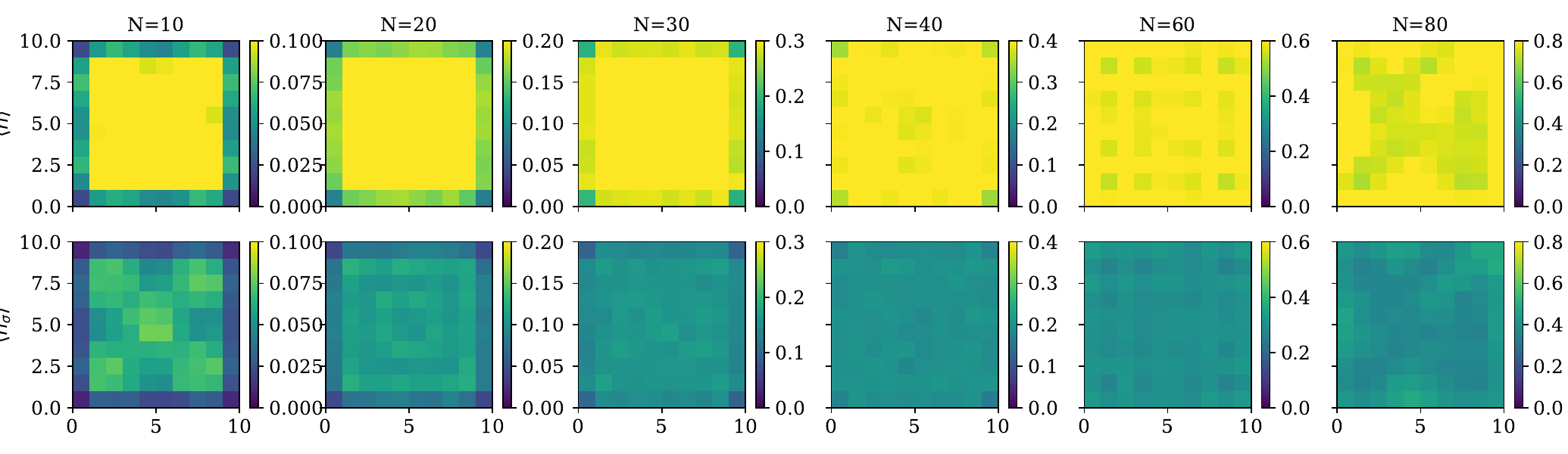}
    \caption{Real space densities $n_i$ per site. The colors are normalized between 0 and $1-\delta=N/L$ for each filling.}
    \label{fig:SM_density}
\end{figure}

In addition to the momentum space distribution, we also show the average real space density $\langle \hat{n}_i\rangle=\sum_\sigma\langle\hat{n}_{i\sigma}\rangle$ as well as the average density per spin $\langle\hat{n}_{i\sigma}\rangle$ in Fig.~\ref{fig:SM_density}. Colorbars are normalized between $0$ and $N/L$ for each doping value.

\section{Density Matrix Renormalization Group Benchmarks \label{sec:DMRG}}
The benchmarks using matrix product states (MPS) shown in this work were obtained using the density matrix renormalization group (DMRG) algorithm \cite{SCHOLLWOCK201196}. Specifically, we use the  the SyTen toolkit \cite{syten1,syten2} with implemented
global $U(1)_N$ and $SU(2)$ symmetries and bond dimensions up to $\chi=2048$ (corresponding to $\approx 12000$ $U(1)$ states). We use the single-site (DMRG3S) in all stages except for the last stage, we the two-site (2DMRG) update scheme is applied. In total, $5$ stages with $40$ sweeps in the first two stages, $20$ sweeps in the intermediate stage and $10$ sweeps in the final stage were used. \\

As discussed in the main text, we do not expect that our MPS are fully converged in the intermediate doping regime. In order to estimate the MPS error, we follow Ref. \cite{Corboz2016} and calculate the truncation error per site $\epsilon_\mathrm{Trunc}$ at each bond dimension. Then, we fit the MPS energies $E_\mathrm{MPS}^\chi$ with a linear dependence on the truncation error, as shown in Fig. \ref{fig:ErrorEstimates} on the right. The lines on the right plot show the respective fits $E_\mathrm{fit}(\epsilon_\mathrm{Trunc})$ for exemplary doping values $\delta$ (the same as shown in Fig. 2). The error is estimated as 
\begin{align}
    \Delta = \left[(E_\mathrm{MPS}^{\chi=2048}-E_\mathrm{fit}(\epsilon_\mathrm{Trunc}=0)\right]/2
\end{align} 
and shown on the left of Fig. \ref{fig:ErrorEstimates}. We find a particularly high error for $\delta=0.4$.

\section{Comparison to previous neural quantum state constructions}

\subsection{Comparison of Gutzwiller-Projected and Unprojected HFDS States}
\begin{figure}[t]
\centering
\includegraphics[width=0.95\textwidth]{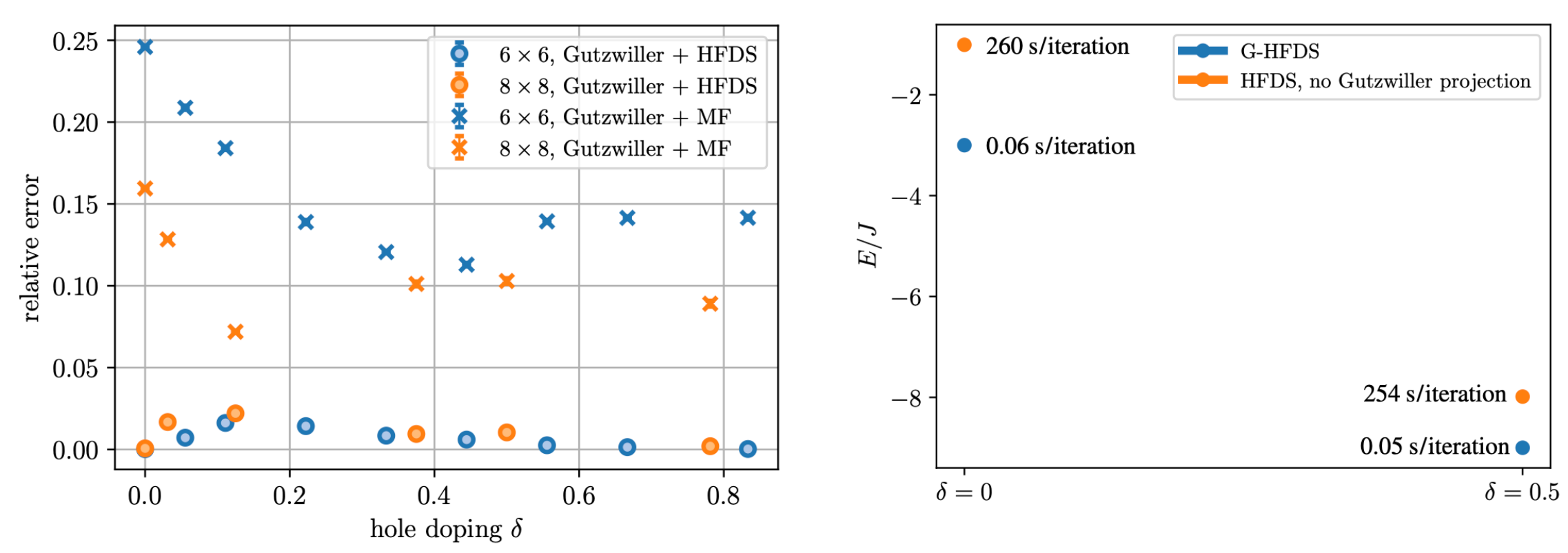}
\caption{Comparison of traditional Gutzwiller ansätze based on a mean-field (G-MF), G-HFDS and HFDS: Left: Relative ground state energy error w.r.t. MPS energies, obtained from G-MF (crosses) versus G-HFDS states (circles) on $6\times6$ and $8\times8$ lattices. Right: Ground state energies for a $2\times 2$ system with doping $\delta=0.0, 0.5$ using G-HFDS (blue) and HFDS without Gutzwiller projection on the level of the ansatz but on the level of teh Hamiltonian. We show the corresponding runtimes as well.}
\label{fig:Gutzwiller}
\end{figure}

In this section, we present a detailed comparison between the Gutzwiller-projected Hidden Fermion Determinant State (G-HFDS) and the original HFDS, to quantify the benefit of incorporating the Gutzwiller projection at the level of the variational ansatz.\\

Figure~\ref{fig:Gutzwiller} (left) shows a comparison of relative ground state energy errors with respect to MPS benchmark results for $6 \times 6$ and $8 \times 8$ systems at various doping levels. The G-HFDS (circles) significantly outperforms the Gutzwiller-projected mean-field states (G-MF, crosses), particularly on larger lattices and at moderate doping. This highlights the added expressivity and accuracy of our approach compared to a MF ansatz.\\

To further emphasize the practical advantages of the G-HFDS ansatz for simulating the $t$-$J$ model, we also compare against a setup where the HFDS is used without the Gutzwiller projection at the level of the ansatz, but instead the projection is applied explicitly at the level of the Hamiltonian. This comparison is shown in the right panel of Fig.~\ref{fig:Gutzwiller} for a $2 \times 2$ system with dopings $\delta = 0.0$ and $0.5$. The unprojected HFDS suffers from two severe limitations:

\begin{itemize}
    \item \textbf{Memory limitations:} Constructing the full projector $\mathcal{P}_G$ involves a product over all sites, which becomes intractable for even moderately sized systems ($4 \times 4$) when using standard hardware such as a single H100 GPU in NetKet.
    
    \item \textbf{Excessive runtimes:} Applying $\mathcal{P}_G$ explicitly results in runtimes that are approximately four orders of magnitude longer. For instance, for the $2 \times 2$ system with four particles, the iteration time increases from roughly $0.06$ seconds (G-HFDS) to $260$ seconds (HFDS with explicit projection). After one hour, the G-HFDS completes over 3000 optimization steps, while the unprojected HFDS manages only 13, leading to much higher final energies.
\end{itemize}

These findings confirm that the Gutzwiller projection is not just an incremental improvement -- it is essential for making the HFDS viable for simulating the $t$–$J$ model. 

\subsection{Comparison to Moreno et al. -- the Fermi-Hubbard model \label{sec:FH}}
\begin{figure}[t]
\centering
\includegraphics[width=0.9\textwidth]{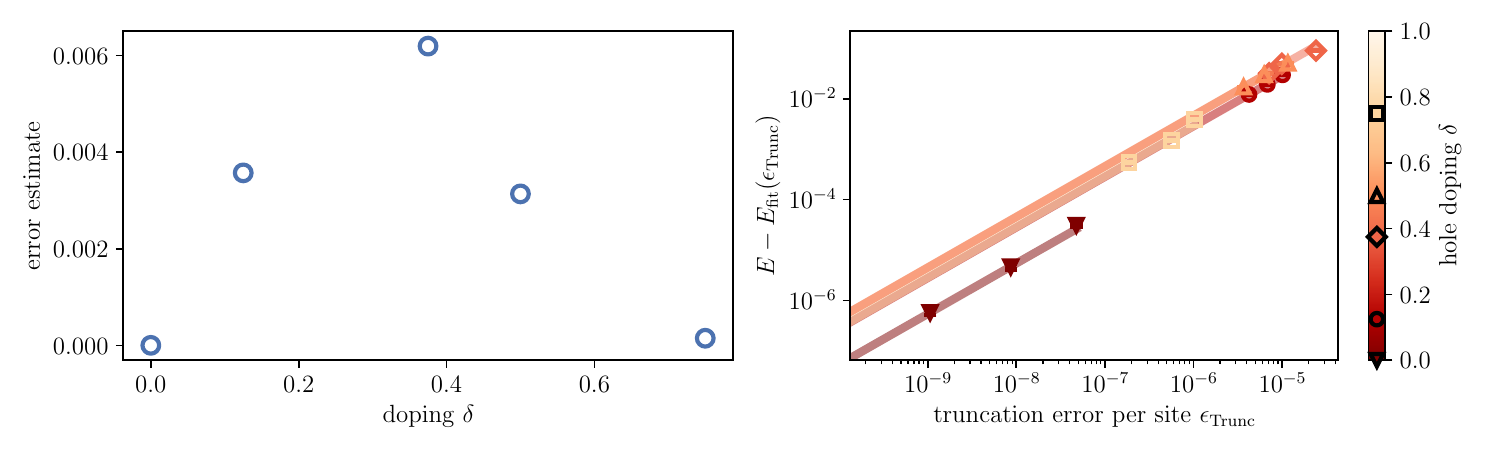}
\caption{Estimation of MPS errors for the $8 \times 8$ systems. Left: The estimated errors on the MPS energies $E_\mathrm{MPS}^\chi$ obtained with bond dimension $\chi=2048$ and both $U(1)$ and $SU(2)$ symmetries. The error estimates are obtained by fitting the MPS energies $E_\mathrm{MPS}^\chi$ w.r.t. the truncation error per site, $\epsilon_\mathrm{Trunc}$. Hereby, the two lowest bond dimensions were neglected. The lines on the right plot show these fits $E_\mathrm{fit}(\epsilon_\mathrm{Trunc})$. The error is estimated as $\left[(E_\mathrm{MPS}^{\chi=2048}-E_\mathrm{fit}(\epsilon_\mathrm{Trunc}=0)\right]/2$ \cite{Corboz2016}. }
\label{fig:ErrorEstimates}
\end{figure}

\begin{figure}[htp]
\centering
\includegraphics[width=0.8\textwidth]{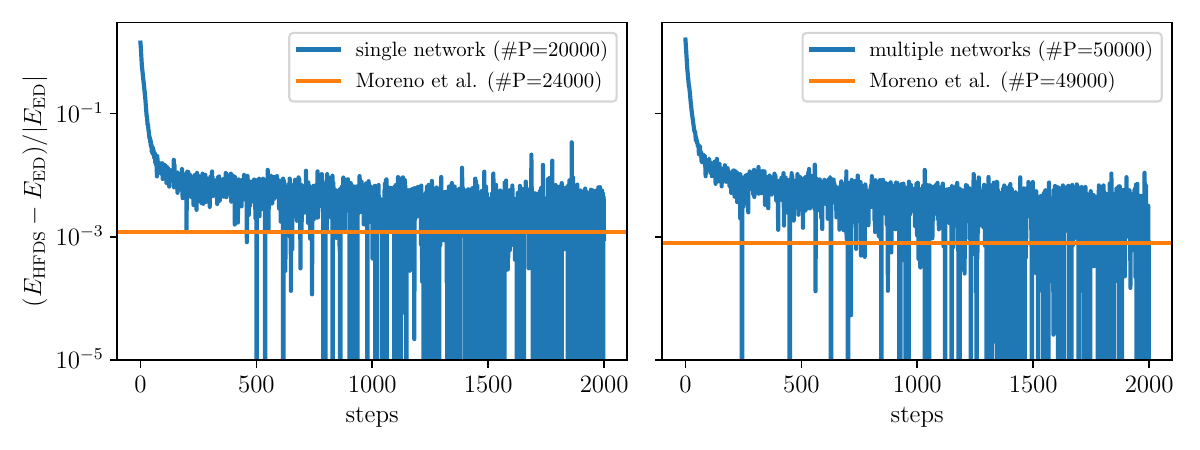}
\caption{Results for the Fermi-Hubbard model on a $4\times 4$ lattice with $N=8$ fermions and periodic boundary conditions. We use a single FFNN network for all entries of the lower block of the Slater determinant (left). This is slightly different to the HFDS used in Moreno et al. \cite{Moreno2022}, where multiple networks for each row of the lower block were employed (right). Hence, the number of parameters \#$P$ differs. We compare our results (blue) with results from Moreno et al. \cite{Moreno2022} (orange) that were obtained with similar \#$P$. In all cases, $8$ hidden fermions were used.}
\label{fig:FHbenchmarks}
\end{figure}

As a first benchmark, we compare the results obtained with our HFDS implementation with results from Moreno et al. \cite{Moreno2022} for the Fermi-Hubbard model on a $4\times 4$ lattice with periodic boundary conditions. In contrast to this work, we use a single FFNN network for all entries of the lower block of the Slater determinant shown in Fig. 1a. This differs slightly from the HFDS used in Moreno et al. \cite{Moreno2022}, where multiple networks for each row of the lower block were employed. In Fig. \ref{fig:FHbenchmarks}, we compare the relative energies obtained with both HFDS variants for $N=8$ fermions on a $4\times 4$ lattice (blue lines). In both cases, we use $8$ hidden fermions and $128$ features. The results agree within the errorbars, although the number of parameters \#$P$ is smaller for the single network. Furthermore, we compare our results (blue) with results from Moreno et al. \cite{Moreno2022} (orange) that were obtained with similar \#$P$. Again, we find a good agreement.

\subsection{Backflow vs. Hidden fermions \label{sec:BackflowVsHidden}}

\begin{figure}[htp]
    \centering
    \includegraphics[width=0.5\textwidth]{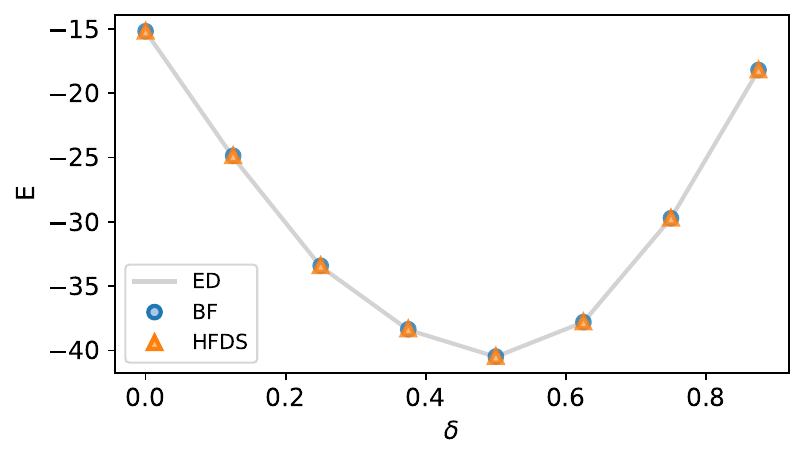}
    \caption{Results for optimized neural backflow states and HFDS. Both states are parametrized by a single FFNN and start from the same random initial state. The network sizes were chosen such that at each doping the number of parameters is approximately equal, between \#$P=22000$ at $\delta=0.875$ and \#$P=71000$ at $\delta=0$.}
    \label{fig:HFDSvsBF}
\end{figure}

We compare the HFDS to the neural backflow architecture for a system of $L_x\times L_y=4\times4$ lattice sites, where we choose the size of the corresponding networks such that they have an approximately equal number of variational parameters. We use the same number of hidden and visible fermions at each doping for the HFDS and adjust the number of features of the backflow network accordingly. Both networks are parameterized by a FFNN with 3 layers and start from a random initial state. Results for the final variational energy are shown in Fig.~\ref{fig:HFDSvsBF}. We find that both architectures lead to the same final energies across all dopings, consistent with the claim that they can be unified within the neural Jastrow-backflow framework~\cite{Liu_2019}.
 
\subsection{Initializations \label{sec:Initializations}}
\begin{figure}
    \centering
    \includegraphics[width=0.4\linewidth]{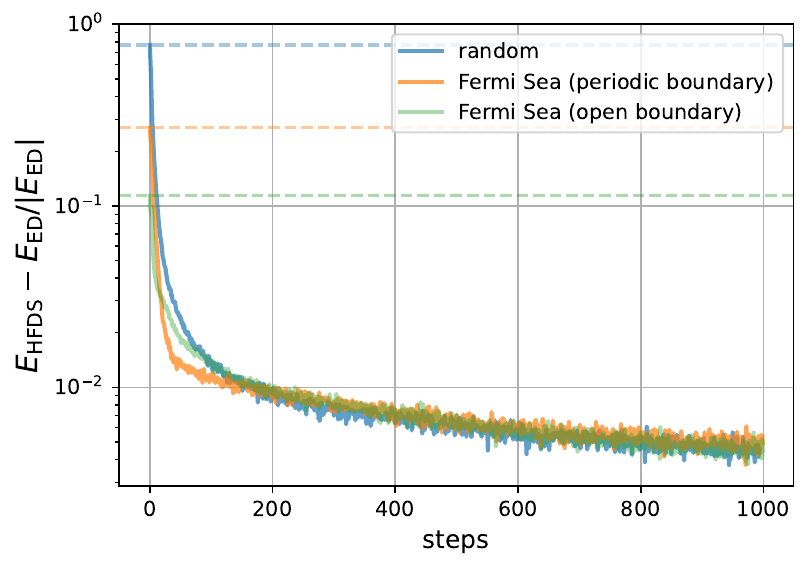}
    \includegraphics[width=0.4\linewidth]{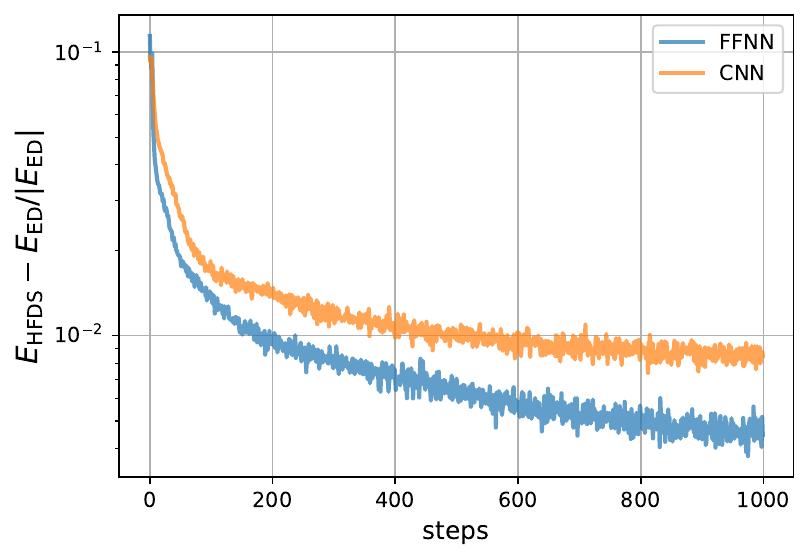}
    \caption{Optimization curves for different initial mean field states (left) and architectures (right) for $\delta=0.125$. All states were optimized with the same network with \#$P=60000$ parameters. Only the initial configuration of the upper left block in the Slater determinant, and the network representing the lower half respectively, was changed. The different architectures are compared for an open boundary Fermi sea initialization. The random state optimization corresponds to the final value at $\delta=0.125$ in Fig.~\ref{fig:HFDSvsBF}.}
    \label{fig:inits}
\end{figure}

We further compare different initial states for the mean field of visible fermions. As described in the main body, the upper left block matrix can be initialized by a physically motivated state, while the other blocks are set to zero or identity, respectively. Fig.~\ref{fig:inits} compares a Gutzwiller projected Fermi sea for both open and periodic boundary conditions to a random parameter initialization. 
The Fermi sea initialization is further described below, see Sec. \ref{sec:FS}. We choose the particularly challenging low doping case $\delta=1/8$ for our benchmarks. 
The Fermi sea is initialized by building the mean field matrix out of the single particle eigenstates of a tight-binding Hamiltonian with either open or periodic boundary conditions. While both Fermi seas achieve similar final energies as the random initial state, we see a slight improvement in convergence time for the Fermi sea states.

\subsubsection*{Fermi sea initialization \label{sec:FS}}

\noindent\textbf{Periodic boundaries:} If we initialize a Fermi sea of a periodic boundary system, the parameters of the momentum eigenstates are obtained straightforwardly by applying a Fourier transform to the real space states \( |\psi_k \rangle = \psi_k(x) |x \rangle \):
\[
\psi_k(x) = e^{ikx} \psi_x.
\]
Note that this makes the mean field parameters complex. Hence, if the rest of the network is real, this leads to a HFDS that has both imaginary and real weights, and complicates the gradient calculation using SR. These complications can in principle be overcome, however, it is easier to circumvent this issue by a basis change: For the free electron model \( E(k) = E(-k) \), so instead of building the Fermi sea out of states \( |k \rangle \) and \( |-k \rangle \) one can use \( |k^+ \rangle = \frac{1}{2} (|k \rangle + |-k \rangle) \) and \( |k^- \rangle = \frac{1}{2i} (|k \rangle - |-k \rangle) \). This leads to new parameters
\[
\psi_{k^+}(x) = \frac{1}{2} (e^{ikx} + e^{-ikx}) \psi_x = \cos(kx) \psi_x,
\]
\[
\psi_{k^-}(x) = \frac{1}{2i} (e^{ikx} - e^{-ikx}) \psi_x = \sin(kx) \psi_x, \tag{A.2}
\]
which are purely real.\\

\noindent\textbf{Open boundaries:} For the open boundary systems discussed in the main text we choose a slightly different initialization by making use of an exact solution of the open boundary single particle tight-binding model,
\[
\hat{H} = \sum_{i,j} \left( \hat{c}_i^{\dagger} \hat{c}_j + \text{h.c.} \right).
\]
Diagonalization gives the single particle orbitals for non-interacting particles in a system with open boundaries.

\subsection{Different network architectures \label{sec:Architectures}}
We further compare the optimization for different network architectures, namely a FFNN and a CNN. Fig.~\ref{fig:inits} shows the relative error of the energy compared to an exact solution. We see that the FFNN achieves slightly better variational energies than the CNN. Both networks are initialized with a projected open boundary Fermi sea.



\end{document}